\documentclass{JHEP3}

\usepackage{latexsym}
\usepackage{graphics}

\def\comment #1{}
\def\cf {{\it cf. }}

\def\refer #1{{(\ref{#1})}}
\def\fullref #1{\ref{#1} (p.\pageref{#1})}

\def\ket #1{\left|{#1}\right\rangle}

\def\of #1{\!\left({#1}\right)}

\def\brackets #1{\left[{#1}\right]}
\def\braces #1{\left\lbrace{#1}\right\rbrace}

\def\commutator #1#2{\brackets{{#1},{#2}}}
\def\antiCommutator #1#2{\braces{{#1},{#2}}}

\def\defas {:=}
\def\shallbe {\stackrel{!}{=}}
\def\equalby #1{\stackrel{\refer{#1}}{=}}

\def\extd {\mathbf{{d}}}

\def\endofproof {$\Box$\newline}

\def\eigenspace #1#2 {\mathrm{eig}\of{#1,#2}}

\newlength{\skiplength}

\def\inner {\!\cdot\!}

\title{Super-Pohlmeyer invariants and boundary states for non-abelian gauge fields}
\author{Urs Schreiber \\ Universit{\"a}t Duisburg-Essen \\ Essen, 45117, Germany\\
   E-mail: \email{Urs.Schreiber@uni-essen.de}}

\abstract{
  Aspects of the supersymmetric extension of the Pohlmeyer invariants are studied,
  and their relation to superstring boundary states for non-abelian gauge fields is discussed.
 
We show that acting with a super-Pohlmeyer invariant with respect to some non-abelian gauge field $A$
on the boundary state of a bare D9 brane produces the boundary state describing that
non-abelian background gauge field on the brane.
Known consistency conditions on that boundary state equivalent to the background equations of
motion for $A$ hence also apply to the quantized Pohlmeyer invariants.
}

\begin{document}

\section{Introduction}

This paper demonstrates a relation between two apparently unrelated 
aspects of superstrings: boundary states for
nonabelian gauge fields and (super-)Pohlmeyer invariants. 

On the one hand side superstring boundary states describing excitations of non-abelian
gauge fields on D-branes are still the subject of investigations 
\cite{MaedaNakatsuOonishi:2004,MurakaniNakatsu:2002,Schreiber:2004e} and
are of general interest for superstring theory, as they directly mediate between
string theory and gauge theory.

On the other hand, studies of string quantization focusing on non-standard 
worldsheet invariants, the so-called Pohlmeyer invariants, done in 
\cite{Pohlmeyer:2002,MeusburgerRehren:2002,Pohlmeyer:1998,Pohlmeyer:1988} and
recalled in \cite{Thiemann:2004}, were shown in \cite{Schreiber:2004b,Schreiber:2004c}
to be related to the standard quantization of the string by way of the well-known DDF invariants. 
This
raised the question whether the Pohlmeyer invariants are of any genuine interest in (super-)string theory
as commonly understood.

Here it shall be shown that the (super-)Pohlmeyer invariants do indeed play an interesting
role as boundary state deformation operators for non-abelian gauge fields, thus connecting
the above two topics and illuminating aspects of both them.\\

A boundary state is a state in the closed string's Hilbert space constructed in such a way
that inserting the vertex operator of that state in the path integral over the sphere reproduces
the disk amplitudes for certain boundary conditions (D-branes) of the open string. In accord with the
general fact that the worldsheet path integral insertions which describe background field excitations
are exponentiations of the corresponding vertex operators, it turns out that the boundary
states which describe gauge field excitations on the D-brane have the form
of (generalized) Wilson lines of the gauge field along the closed string 
\cite{Hashimoto:2000,Hashimoto:1999,MaedaNakatsuOonishi:2004,MurakaniNakatsu:2002,Schreiber:2004e}.

Long before these investigations, it was noted by Pohlmeyer \cite{Pohlmeyer:1988}, in the context of the classical 
string, that generalized Wilson lines along the closed string with respect to an auxiliary
gauge connection on spacetime provide a ``complete'' set of invariants of the theory, i.e.
a complete set of observables which (Poisson-)commute with all the Virasoro constraints.

Given these two developments it is natural to suspect that there is a relation between Pohlmeyer
invariants and boundary states. Just like the DDF invariants (introduced in \cite{DelGiudiceDiVecchiaFubini:1972}
and recently reviewed in \cite{Schreiber:2004b}), 
which are the more commonly considered
complete set of invariants of the string, commute with all the constraints and hence generate physical
states when acting on the worldsheet vacuum, a consistently quantized version of the
Pohlmeyer invariants should send boundary states of bare D-branes to those involving the
excitation of a gauge field.

Indeed, up to a certain condition on the gauge field, this turns out to be true and works as follows:\\

If $X^\mu(\sigma)$ and $P_\mu(\sigma)$ are the canonical coordinates and momenta of the
bosonic string, then
$\mathcal{P}_\pm^\mu\of{\sigma} \defas \frac{1}{\sqrt{2}T}(P_\mu\of{\sigma} \pm T \eta_{\mu\nu}X^{\prime\nu}\of{\sigma})$,
(where $T$ is the string's tension and a prime denotes the derivative with respect to $\sigma$)
are the left- and right-moving bosonic worldsheet fields for flat Minkowski background (in CFT context
denoted by $\partial X$ and $\bar \partial X$) and for any given constant gauge field $A$ on target space
the objects
\begin{eqnarray}
  \label{Pohlmeyer in introduction}
  W_\pm^{\mathcal{P}}[A]
  &\defas&
  \mathrm{Tr}\, \mathrm{P}
  \exp\of{\int_0^{2\pi}d\sigma\, A \inner \mathcal{P}_\pm\of{\sigma}}
\end{eqnarray}
(where $\mathrm{Tr}$ is the trace in the given representation of the 
gauge group's Lie algebra and $\mathrm{P}$ denotes path-ordering
along $\sigma$) Poisson-commute with all Virasoro constraints. In fact 
the coefficients of $\mathrm{Tr}(A^n)$ in these generalized Wilson lines do so seperately, and these
are usually addressed as the \emph{Pohlmeyer invariants}, even though we shall use this term 
for the full object \refer{Pohlmeyer in introduction}.

Fundamentally, the reason for this invariance is just the reparameterization invariance of the Wilson line,
which can be seen to imply that \refer{Pohlmeyer in introduction} remains unchanged under a substitution of
$\mathcal{P}$ with a reparameterized version of this field. In \cite{Schreiber:2004b} it was
observed that an interesting example for such a substitution is obtained by taking the ordinary DDF oscillators
\begin{eqnarray}
  \label{bosonic DDF in intro}
  A_m^\mu 
  &\propto&
  \int\limits_0^{2\pi}
  d\sigma\;
  \mathcal{P}_-^\mu\of{\sigma}e^{im\frac{4 \pi T}{k\inner p}\, k\inner X_-\of{\sigma}}
\end{eqnarray}
(where $k$ is a lightlike vector on target space,
$X_-$ is the left-moving component of $X$, $p$ is the center-of-mass momentum,  and an analogous expression
exists for $\mathcal{P}_+$)
and forming ``quasi-local'' invariants
\begin{eqnarray}
  \label{bosonic quasi local in intro}
 \mathcal{P}_-^{R\mu}\of{\sigma}
  &\defas&
  \frac{1}{\sqrt{2\pi}}
  \sum\limits_{m=0}^\infty
  A_m^\mu e^{im\sigma}
\end{eqnarray}
from them.\footnote{We dare to use the same symbol $A$ for the gauge field and for the DDF oscillators
in order to comply with established conventions. The DDF oscillators will always carry a mode index
$m$, however, and it should always be clear which object is meant.} 

One finds 
\begin{eqnarray}
  W^{\mathcal{P}}[A] &=& W^{\mathcal{P}^R}[A]
\end{eqnarray}
 and since the quantization of the $\mathcal{P}^R$
in terms of DDF oscillators is well known, this gives a consistent quantization of the Pohlmeyer invariants.
This is the quantization that we shall use here to study boundary states.

The above construction has a straightforward generalization to the superstring and this is the context
in which the relation between the Pohlmeyer invariants and boundary states turns out to have interesting aspects,
(while the bosonic case follows as a simple restriction, when all fermions are set to 0).

So we consider the supersymmetric extension of \refer{bosonic DDF in intro}, which, by 
convenient abuse of notation,
we shall also denote by $A_m^\mu$:
\begin{eqnarray}
  \label{susy DDF in intro}
  A_m^\mu
  &\propto&
  \int_0^{2\pi}
  d\sigma
  \left(
    \mathcal{P}_-\of{\sigma}
    +im \frac{\pi \sqrt{2T}}{k\inner p}
    k\inner\Gamma_-\of{\sigma}\, \Gamma_-^\mu\of{\sigma},
    e^{im\frac{4 \pi T}{k\inner p}\, k\inner X_-\of{\sigma}}
  \right)
  \,,
\end{eqnarray}
where $\Gamma_\pm\of{\sigma}$ denote the fermionic superpartners of $\mathcal{P}_\pm$. From
these we build again the objects \refer{bosonic quasi local in intro} and 
finally $W^{\mathcal{P}^R}[A]$, which we address as the \emph{super-Pohlmeyer} invariants. 

Being constructed from the supersymmetric invariants $\mathcal{P}^R$, which again are 
built from \refer{susy DDF in intro}, 
these manifestly commute with all of the super-Virasoro constraints. But in order to
relate them to boundary states they need to be re-expressed in terms of the plain objects
$\mathcal{P}$ and $\Gamma$. This turns out to be non-trivial and has some interesting aspects
to it. \\

After these peliminaries we can state 
the first result to be reported here, which is
\begin{enumerate}
\item
that on that subspace $\mathbf{P}_k$ of phase space where $k\inner X_-$
is invertible as a function of $\sigma$ (a condition that plays also a crucial role for the 
considerations of the bosonic DDF/Pohlmeyer relationship as discussed in \cite{Schreiber:2004b})
the super-Pohlmeyer invariants built from \refer{susy DDF in intro} are equal to 
\begin{eqnarray}
  \label{restricted super-Pohlmeyer invariant in intro}
  \left.
    W^{\mathcal{P}^R}[A]
  \right|_{\mathbf{P}_k}
  &=&
  \mathrm{Tr}\;
  \mathrm{P}
  \exp\of{
    \int_0^{2\pi}
    d\sigma\;
    \left(
      i A_\mu + 
      \commutator{A_\mu}{A_\nu}
      \frac{
k \inner \Gamma\;
      \Gamma^\nu
}{2k\inner \mathcal{P}}
    \right)
    \mathcal{P}^\mu
  }
  \,,
\end{eqnarray}

\item 
 that this expression extends to an invariant on all of phase space precisely if
the transversal components of $A$ mutually commute,

\item
  and that in this case the above is equal to
\begin{eqnarray}
  \label{nonlocal susy extension of Pohlmeyer intro}
  Y[A]
  &\defas&
  \mathrm{Tr}\,\mathrm{P}
  \exp\of{\int_0^{2\pi}d\sigma\;
    \left(
    i A_\mu \mathcal{P}^\mu
    +
    \frac{1}{4}\commutator{A_\mu}{A_\nu}
    \Gamma^\mu \Gamma^\nu
    \right)
  }
  \,.
\end{eqnarray}

\end{enumerate}

The second result concerns the application of the quantum version of these observables to 
the bare boundary state $\ket{\mathrm{D9}}$ of a space-filling D9-brane 
(see for instance appendix A of \cite{Schreiber:2004e} for a brief review of boundary state formalism and
further literature). Denoting by
$\mathcal{E}^\dagger\of{\sigma} = \frac{1}{2}\left(\Gamma_+\of{\sigma} + \Gamma_-\of{\sigma}\right)$
the differential forms on loop space (\cf section 2.3.1. of \cite{Schreiber:2004e} and
section 2.2 of \cite{Schreiber:2004} for the notation and nomenclature used here, and see
\cite{Schreiber:2004f} for a more general discussion of the loop space perspective)
we find
  that for the above case of commuting transversal $A$ the application of 
\refer{nonlocal susy extension of Pohlmeyer intro} to $\ket{\mathrm{D9}}$ yields
\begin{eqnarray}
  &&\mathrm{Tr}\, \mathrm{P}
  \exp\of{
    \int\limits_0^{2\pi}
     d\sigma\,
    \left(
      i A_\mu \mathcal{P}^\mu + \frac{1}{4}(F_A)_{\mu\nu}\Gamma^\mu\Gamma^\nu
    \right)
  }
  \ket{\mathrm{D9}}
  \nonumber\\
  &=&
  \mathrm{Tr}\, \mathrm{P}
  \exp\of{
    \int\limits_0^{2\pi}
     d\sigma\,
    \left(
      i 
      A_\mu X^{\prime \mu} + \frac{1}{4}(F_A)_{\mu\nu}
      \mathcal{E}^{\dagger \mu}
      \mathcal{E}^{\dagger \nu}
    \right)
  }
  \ket{\mathrm{D9}}
  \,.
\end{eqnarray}
which is, on the right hand side, precisely the boundary state describing a non-abelian gauge field
on the D9 brane \cite{MaedaNakatsuOonishi:2004,Schreiber:2004e} .

In summary this shows that and under which conditions the application of a quantized super-Pohlmeyer
invariant to the boundary state of a bare D9 brane produces the boundary state describing a
non-abelian gauge field excitation. \\

The structure of this paper closely follows the above outline:

First of all \S\fullref{Super-Pohlmeyer invariants} 
is concerned with the classical super-Pohlmeyer invariants and
their expression in terms of local fields. Then \S\fullref{Super-Pohlmeyer invariants} 
discusses their cousins, the
invariants of the general form \refer{nonlocal susy extension of Pohlmeyer intro}.
Both are related in \S\fullref{Invariance of the extension of the restricted super-Pohlmeyer invariants}. 

Then the quantization of the super-Pohlmeyer invariants
is started in \S\fullref{Quantum super-Pohlmeyer invariants}. 
After an intermediate result concerning an operator ordering issue
is treated in \S\fullref{On an operator ordering issue in Wilson lines along the closed string} 
the quantum Pohlmeyer invariants are finally 
applied to the bare boundary state in \S\fullref{Super-Pohlmeyer and boundary states}.

\S\fullref{Summary and Conclusion} gives some concluding remarks.

\section{DDF operators, Pohlmeyer invariants and boundary states}

\subsection{Super-Pohlmeyer invariants}
\label{Super-Pohlmeyer invariants}

In \cite{Schreiber:2004b} it was shown how from the classical DDF oscillators of the bosonic string one can construct
\emph{quasilocal} fields $\mathcal{P}^R$, which (Poisson-)commute with all the constraints and
which, when used in place of $X^\prime$ in a Wilson line of a constant gauge field
along the string, reproduce the Pohlmeyer invariants. It was mentioned that using the DDF oscillators
of the superstring in this procedure leads to a generalization of the Pohlmeyer invariants
to the superstring. Here we will work out the explicit form of the \emph{super-Pohlmeyer invariants}
obtained this way and point out that they are interesting in their own right.

Using the notation of \cite{Schreiber:2004b} we denote by $\mathcal{P}^\mu(\sigma)$ the 
classical canonical left- or right-moving bosonic fields on the string, and by $\Gamma^\mu(\sigma)$
their fermionic partners, where the relation to the usual CFT notation is 
$\mathcal{P}^\mu \propto \partial X^\mu$
and $\Gamma^\mu \propto \psi^\mu$.

Our normalization is chosen such that the graded Poisson-brackets read
\begin{eqnarray}
  \label{classical CCM}
  &&\commutator{\Gamma^\mu\of{\sigma}}{\Gamma^\nu\of{\kappa}}
  =
  -2\eta^{\mu\nu}\delta\of{\sigma-\kappa}
  \nonumber\\
  &&
  \commutator{\mathcal{P}^\mu\of{\sigma}}{\mathcal{P}^\nu\of{\kappa}}
  =
  -\eta^{\mu\nu}\delta^\prime\of{\sigma-\kappa}
  \,.
\end{eqnarray}

The classical bosonic DDF oscillators $A_m^\mu$ of the superstring are obtained by acting with the supercharge
\begin{eqnarray}
  \label{the supercharge}
  G_0 
  &=&
  \frac{i}{\sqrt{2}}
  \int
  d\sigma\;
  \Gamma^\mu \mathcal{P}_\mu 
\end{eqnarray}
(we concentrate on the Ramond sector for notational simplicity) on integrals over weight 1/2 fields:
\begin{eqnarray}
  \label{bosonic susy DDF oscillator}
  A_m^\mu
  &\defas&
  \commutator{
    G_0  
  }{
    \frac{i}{\sqrt{4\pi}}
    \oint d\sigma\;
    \Gamma^\mu e^{-imR}
}
  \nonumber\\
  &=&
  \frac{1}{\sqrt{2\pi}}
  \oint d\sigma\;
  \left(
    {\cal P}^\mu
    +
    i
    m
    \frac{\pi\sqrt{2T}}{k \inner p}
    k \inner \Gamma
    \Gamma^\mu
  \right)
  e^{-im R}
  \,,
\end{eqnarray}
where
\begin{eqnarray}
    R\of{\sigma}
  &\defas&
  -
  \frac{4\pi T}{k\inner p}\,
    k\inner X_\pm\of{\sigma}
\end{eqnarray}
and $p^\mu = \int_0^{2\pi}P^\mu\of{\sigma}$.

By construction, the $A_m^\mu$ super-Poisson-commute with all the constraints.
From the $A_m^\mu$ quasi-local objects $\mathcal{P}^R$ are reobtained by Fourier transforming from the integral
mode index $m$ to the parameter $\sigma$:

\begin{eqnarray}
  \label{quasilocal observable}
  \mathcal{P}^{R}\of{\sigma}
  &\defas&
  \frac{1}{\sqrt{2\pi}}
  \sum\limits_{n = -\infty}^\infty
  A_n\,
  e^{in\sigma}
  \nonumber\\
  &=&
  \int_0^{2\pi}
  d\tilde \sigma\;
  \left(
    \mathcal{P}\of{\tilde \sigma}
    \delta\of{R\of{\tilde\sigma} - \sigma}
    +
    \frac{i\pi \sqrt{2T}}{k\inner p}
    k\inner \Gamma\of{\tilde \sigma}
    \Gamma\of{\tilde \sigma}
    \frac{\partial}{\partial \sigma}\delta\of{R\of{\tilde \sigma}-\sigma}
  \right)
  \,.
\end{eqnarray}

The role of the DDF oscillators played here is the derivation of this expression. Their invariance
was rather easy to enforce and check, but by taking combinations of them as in
\refer{quasilocal observable} and further constructions below, 
we can now build objects which are necessarily still invariants,
but whose invariance is much less obvious.

Since the DDF oscillators $A_m^\mu$ won't be explicitly needed anymore in the following, we
take the liberty to reserve the letter $A$ from now on to describe a gauge connection on target space.
We shall be interested in the Wilson line
\begin{eqnarray}
  \label{generalized Wilso line}
  W^{\mathcal{P}^R}[A]
  &\defas&
  \mathrm{Tr}\,
  \mathrm{P}
  \exp\of{
    i
    \int_0^{2\pi}
    d\sigma\;
    A\inner \mathcal{P}^{R}\of{\sigma}
  }
\end{eqnarray}
with respect to this gauge connection $A$, constructed using the ``generalized tangent vector'' $\mathcal{P}^{R}$
which plays the role of the true tangent vector $X^\prime$ found in ordinary Wilson lines.
Because this object follows in spirit closely the construction principle of the bosonic Pohlmeyer invariants,
and because its bosonic component coincides with the purely bosonic Pohlmeyer invariant, we shall here
address it as the \emph{super-Pohlmeyer invariant}. 	In the following a
form of this object in terms of the original local fields $\mathcal{P}$ and $\Gamma$ is derived, 
which will illuminate its relation to supersymmetric boundary states.

The integrand of \refer{generalized Wilso line} can be put in a more insightful form by means of 
a couple of manipulations:

Following the development in \cite{Schreiber:2004b} (\cf equation (2.43)) 
we now temporarily restrict attention to the subspace
$\mathbf{P}_k$
of phase space on which the function $R$ is invertible, in which case it is, by construction,
$2\pi$-periodic.  On that part of phase space (and only there)
the integral in \refer{quasilocal observable} can be evaluated to yield
\begin{eqnarray}
  \label{quasi local super P}
  \left.\mathcal{P}^R\of{\sigma}\right|_{\mathbf{P}_k}
  &=&
  \left(
    R^{-1}
  \right)^\prime
  \of{\sigma}
  \mathcal{P}
  \of{
    R^{-1}\of{\sigma}  
  }
  +
  \frac{\pi i \sqrt{2T}}{k \inner p}
  \frac{\partial}{\partial\sigma}
  \left(
  \left(
    R^{-1}
  \right)^\prime
  \of{\sigma}
  k \inner \Gamma\of{R^{-1}\of{\sigma}}
  \Gamma\of{R^{-1}\of{\sigma}}
  \right)
  \,.
  \nonumber\\
\end{eqnarray}
The first term is known from the bosonic theory (equation (2.51) in \cite{Schreiber:2004b}). The second
term involves the fermionic correction due to supersymmetry, and its remarkable property is
that it is a total $\sigma$-derivative. This means that when $\mathcal{P}^R$ is inserted in
a multi-integral as they appear in \refer{generalized Wilso line}, the fermionic term will produce boundary terms
and hence coalesce with neighbouring integrands. 

Before writing this down in more detail first note that due to $k$ being a null vector the
fermionic terms can never coalesce with themselves, because of
\begin{eqnarray}
  \label{some vanishing of fermionic combinations}
  &&
  \left(
    R^{-1}
  \right)^\prime
  k \inner \Gamma\of{R^{-1}}
  A \inner \Gamma\of{R^{-1}}
  \of{\sigma}
 \frac{\partial}{\partial\sigma}
  \left(
  \left(
    R^{-1}
  \right)^\prime
  k \inner \Gamma\of{R^{-1}}
  A \inner \Gamma\of{R^{-1}}
  \right)
  \of{\sigma}
  \nonumber\\
  &=&
  \frac{1}{2}
  \frac{\partial}{\partial \sigma}
  \underbrace{
  \left(
  \left(
    R^{-1}
  \right)^\prime
  k \inner \Gamma\of{R^{-1}}
  A_\mu\Gamma^\mu_-\of{R^{-1}}    
  \right)^2\of{\sigma}}_{ = 0}
  \nonumber\\
  &=&
  0
  \,.
\end{eqnarray}
This vanishing result depends on the Grassmann properties of the classical fermions
$\Gamma$, which we are dealing with here. 
The generalization of the present development to the quantum theory requires
more care and is dealt with below.

Using \refer{some vanishing of fermionic combinations} a little reflection shows that, when the total derivative 
terms in \refer{generalized Wilso line} are all integrated over and coalesced at the integration bounds
with the neighbouring terms $i A\inner \mathcal{P}$, this yields

\hspace{-1.5cm}\parbox{12cm}{
\begin{eqnarray}
  \label{super-Pohlmeyer invariant, ugly form}
  &&\left.\mathrm{Tr}\;
  \mathrm{P}
  \exp\of{i
    \int_0^{2\pi}
    d\sigma\;
    A \inner \mathcal{P}^R\of{\sigma}
  }\right|_{\mathbf{P}_k}
  \nonumber\\
  &=&
  \mathrm{Tr}\;
  \mathrm{P}
  \exp\of{
    \int_0^{2\pi}
    d\sigma\;
    \left(
      i A_\mu + 
      \commutator{A_\mu}{A_\nu}
      \frac{\pi \sqrt{2T}}{k\inner p}
      (R^{-1})^\prime\of{\sigma}
      k \inner \Gamma\of{R^{-1}\of{\sigma}}\;
      \Gamma^\nu\of{R^{-1}\of{\sigma}}
    \right)
    (R^{-1})^\prime\of{\sigma}
    \mathcal{P}^\mu\of{R^{-1}\of{\sigma}}
  }
  \,.
  \nonumber\\
\end{eqnarray}
}

This expression simplifies drastically when a change of variable $\tilde \sigma \defas R^{-1}\of{\sigma}$ is
performed in the integral, as in (2.23) of \cite{Schreiber:2004b}:
\begin{eqnarray}
  \cdots 
  &=&
  \mathrm{Tr}\;
  \mathrm{P}
  \exp\of{
    \int_0^{2\pi}
    d\tilde \sigma\;
    \left(
      i A_\mu + 
      \commutator{A_\mu}{A_\nu}
      \frac{\pi  \sqrt{2T}}{k\inner p}
      (R^{-1})^\prime\of{R\of{\tilde \sigma}}
      k \inner \Gamma\of{\tilde \sigma}\;
      \Gamma^\nu\of{\tilde \sigma}
    \right)
    \mathcal{P}^\mu\of{\tilde \sigma}
  }
  \,.
  \nonumber\\
\end{eqnarray}
The fermionic term further simplifies by using $(R^{-1})^\prime\of{R\of{\tilde \sigma}} = 
1/R^\prime\of{\tilde \sigma}$ and then equation (2.42) of \cite{Schreiber:2004b}, which gives
$(R^{-1})^\prime\of{R\of{\tilde \sigma}} = \frac{k\inner p}{2\pi \sqrt{2T}}\frac{1}{k\inner \mathcal{P}}$.
This way the above is finally rewritten as
\begin{eqnarray}
  \label{super-Pohlmeyer invariant}
  \left.
    W^{\mathcal{P}^R}[A]
  \right|_{\mathbf{P}_k}
  &=&
  \mathrm{Tr}\;
  \mathrm{P}
  \exp\of{
    \int_0^{2\pi}
    d\sigma\;
    \left(
      i A_\mu + 
      \commutator{A_\mu}{A_\nu}
      \frac{
k \inner \Gamma\;
      \Gamma^\nu
}{2k\inner \mathcal{P}}
    \right)
    \mathcal{P}^\mu
  }
  \,.
  \nonumber\\
\end{eqnarray}
This is the advertized explicit form of the super-Pohlmeyer invariant in terms of local fields,
when restricted to $\mathbf{P}_k$. 

The right hand side extends to an observable on all of phase space in the obvious way
and it is of interest to study if this extension is still an invariant. 
This is the content of the following subsections.

\subsection{Another supersymmetric extension of the bosonic Pohlmeyer invariants}
\label{Another supersymmetric extension of the bosonic Pohlmeyer invariants}

We address the objects
\refer{generalized Wilso line}
as super-Pohlmeyer invariants, because they are obtained from the bosonic Pohlmeyer invariants written in the
form 
$\mathrm{Tr}\,\mathrm{P}\exp\of{\int_0^{2\pi}d\sigma\, A\inner \mathcal{P}^R\of{\sigma}}$
of equation (2.52) of \cite{Schreiber:2004b} by replacing the bosonic \emph{quasi-local} invariants
$\mathcal{P}^R$ by their supersymmetric version \refer{quasilocal observable}. In this sense this
supersymmetric extension is \emph{local}, or rather ``quasi-local'', since the
$\mathcal{P}^R$ are. But it turns out that there is another fermionic extension of the
bosonic Pohlmeyer invariant $\mathrm{Tr}\, \mathrm{P}\exp\of{\int_0^{2\pi} d\sigma\;A \inner \mathcal{P}\of{\sigma}}$
which Poisson-commutes with all the super-Virasoro generators, and which is not local in this sense, namely 
\begin{eqnarray}
  \label{nonlocal susy extension of Pohlmeyer}
  Y[A]
  &\defas&
  \mathrm{Tr}\,\mathrm{P}
  \exp\of{\int_0^{2\pi}d\sigma\;
    \left(
    i A_\mu \mathcal{P}^\mu\of{\sigma}
    +
    \frac{1}{4}\commutator{A_\mu}{A_\nu}
    \Gamma^\mu\of{\sigma} \Gamma^\nu\of{\sigma}
    \right)
  }
  \,.
\end{eqnarray}
Here the integrand itself does not Poisson-commute with the supercharge $G_0$, but $Y[A]$ 
as a whole does.
(This can easily be generalized even to non-constant $A$, but we will here be content with writing down
all expression for the case of constant $A$. Non-constant $A$ will be discussed in the context of the
quantum theory further below.)

Invariance under the bosonic Virasoro generators is  immediate, because the integrand has unit weight.
All that remains to be checked is hence
\begin{eqnarray}
  \commutator{G_0}{Y[A]} &=& 0
  \,.
\end{eqnarray}

{\it Proof:}
This is best seen by following the logic involved in the derivation of equation (3.8) 
in \cite{Schreiber:2004e}:
There are terms coming from
$
  \commutator{G_0}{iA\inner \mathcal{P}\of{\sigma}}
  \propto
  i A\inner \Gamma^\prime\of{\sigma}  
$
which coalesce at the integration boundary with $iA\inner \mathcal{P}$ to give 
$
  -\commutator{A_\mu}{A_\nu} \Gamma^\mu \mathcal{P}^\nu
  \,.
$
This cancels with the contribution from
$
  \commutator{G_0}{\frac{1}{4}\commutator{A_\mu}{A_\nu}\Gamma^\mu\Gamma^\nu}
  \propto
  \commutator{A_\mu}{A_\nu}\Gamma^\mu\mathcal{P}^\nu
  .
$
(Here we write $\propto$ only as a means to ignore the irrelevant global prefactor $i/\sqrt{2}$ in
\refer{the supercharge}.)
Moreover, there is coalescence of $A\inner \Gamma^\prime$ with 
$\commutator{A_\mu}{A_\nu}\Gamma^\mu\Gamma^\nu$ which yields
$\commutator{A_\kappa}{\commutator{A_\mu}{A_\nu}}\Gamma^\kappa\Gamma^\mu\Gamma^\nu = 0$,
so that everything vanishes. This establishes the full invariance of $Y[A]$ under the
super-Virasoro algebra.
\endofproof

With this insight in hand, one can make a curious observation. Write $A_+ \defas k\inner A$
and consider the special case where all transversal components of $A$ mutually commute
\begin{eqnarray}
  \label{transversal A mutually commute}
  \commutator{A_i}{A_j} &=& 0\,,\;\; \forall\, i,j \neq +
  \,.
\end{eqnarray} 
Then
\begin{eqnarray}
  \commutator{A_\mu}{A_\nu}\frac{k\inner \Gamma \Gamma^\nu}{2 k\inner \mathcal{P}} \mathcal{P}^\mu
  &=&
  \frac{1}{2}\commutator{A_+}{A_i}\Gamma^+ \Gamma^i
  \nonumber\\
  &=&
  \frac{1}{4}
  \commutator{A_\mu}{A_\nu}\Gamma^\mu \Gamma^\nu
  \,.
\end{eqnarray}
Comparison of \refer{super-Pohlmeyer invariant} with \refer{nonlocal susy extension of Pohlmeyer} 
hence shows that in this case the super-Pohlmeyer invariant \refer{super-Pohlmeyer invariant} and the
invariant \refer{nonlocal susy extension of Pohlmeyer} coincide:
\begin{eqnarray}
  \label{two Pohlmeyer susy extensions coincide}
  \commutator{A_i}{A_j} = 0\,,\;\; \forall\, i,j \neq +
  &\Rightarrow&
  \left.
    W^{\mathcal{P}^R}[A]
  \right|_{\mathbf{P}_k}
  =
  Y[A]
  \,.
\end{eqnarray}
So in particular in the case \refer{transversal A mutually commute} the extension of the right hand
side of \refer{super-Pohlmeyer invariant} to all of phase space is still an invariant.

Comparing \refer{super-Pohlmeyer invariant} with equation (3.14) of \cite{Schreiber:2004e}
it is obvious, and will be discussed in more detail below, 
that $Y[A]$ must somehow be closely related to the boundary deformation operator describing
non-abelian $A$-field excitations. Together with \refer{two Pohlmeyer susy extensions coincide}
this gives a first indication of how super-Pohlmeyer invariants give insight into boundary states
of the superstring.

Before discussing this in more detail the next section investigates the most general condition
under which the extension of the right hand side of \refer{super-Pohlmeyer invariant} to all of phase space
is still an invariant. It turns out that there are other cases besides \refer{transversal A mutually commute}.

\subsection{Invariance of the extension of the restricted super-Pohlmeyer invariants}
\label{Invariance of the extension of the restricted super-Pohlmeyer invariants}

For the bosonic string the constraint $\mathcal{P}\inner \mathcal{P} = 0$, which
says that $\mathcal{P}$ is a null vector in target space, ensured that
the invertibility of $R$ was preserved by the evolution generated by the constraints
(\cf the discussion on p.12 of \cite{Schreiber:2004b}). 

The same is no longer true for the
superstring, where we schematically have $\mathcal{P}\inner \mathcal{P} + \Gamma\inner \Gamma^\prime = 0$,
instead. It follows that we cannot expect the extension of the right hand side of
\refer{super-Pohlmeyer invariant} to all of (super-)phase space to super-Poisson commute
with all the constraints, since the flow induced by the constraints will in general
leave the subspace $\mathbf{P}_k$. Only for the bosonic string does the flow induced by the
constraints respect $\mathbf{P}_k$.

Notice that this is not in contradiction to the above result that on $\mathbf{P}_k$ the
super-Pohlmeyer invariant \refer{generalized Wilso line} (which by construction super-Poisson commutes
with all the constraints) coincides with \refer{super-Pohlmeyer invariant}. Two functions
which conincide on a subset of their mutual domains need not have coinciding derivatives
at these points.

First of all one notes that the invariance under the action of the bosonic constraints
is still manifest in \refer{super-Pohlmeyer invariant}. Because the integrand still has unit weight
one checks this simply by using the same reasoning as in equation (2.19) of \cite{Schreiber:2004b}.

But the result of super-Poisson commuting with the supercharge $G_0$ is rather non-obvious.
A careful calculation shows that the result vanishes if and only if
\begin{eqnarray}
  \label{first condition  for invariance}
  \commutator{A_i}{A_j} &=& 0\,,\;\;\forall\, i,j\neq +
\end{eqnarray}
or
\begin{eqnarray}
  \label{second condition  for invariance}
  k\inner \Gamma^\prime = 0 = k\inner \mathcal{P}^\prime
  \,.
\end{eqnarray}

The first condition is that already discussed in 
\S\fullref{Another supersymmetric extension of the bosonic Pohlmeyer invariants}.
The second condition is nothing but the defining condition of \emph{lightcone gauge}
on the worldsheet.

Notice that these two conditions are very different in character. When the first
\refer{first condition  for invariance} is satisfied it means that the extension of the
right hand side of \refer{super-Pohlmeyer invariant} to all of phase space is indeed 
an honest invariant. When the first condition is not satisfied then the extenstion of the
right hand side of \refer{first condition  for invariance} to all of phase space
is simply not an invariant. Still, it is an object whose Poisson-commutator with the
super-Virasoro constraints vanishes on that part of phase space where
\refer{second condition  for invariance} holds.

We now conclude this subsection by giving the detailed {\it proof} for the above two conditions.

{\it Proof:}

First consider the
terms of fermionic grade 1. These are contributed by 
\begin{eqnarray}
  \label{an equation}
  \commutator{G_0}{A\inner \mathcal P} \propto A\inner \Gamma^\prime
\end{eqnarray}
as well as 
\begin{eqnarray}
  \label{another equation}
  \commutator{A_\mu}{A_\nu}
\frac{\commutator{G_0}{k\inner \Gamma}\Gamma^\nu \mathcal{P}^\mu}{2 k\inner \mathcal{P}}
  &\propto&
  -
  \commutator{A_\mu}{A_\nu}\Gamma^\nu\mathcal{P}^\mu
  \,.
\end{eqnarray} 
The other remaining
fermionic contraction does not contribute, due to
\begin{eqnarray}
  \commutator{A_\mu}{A_\nu}\commutator{G_0}{\Gamma^\nu}\mathcal{P}^\mu \propto
  -2\commutator{A_\mu}{A_\nu}\mathcal{P}^\nu\mathcal{P}^\mu = 0
  \,.
\end{eqnarray}
In the path ordered integral the terms \refer{an equation} appear as
\begin{eqnarray}
  &&
  \cdots iA\inner \mathcal{P}\of{\sigma_{i-1}}
  \int_{\sigma_{i-1}}^{\sigma_{i+1}}
  iA \inner \Gamma^\prime\of{\sigma_i}
  \; d\sigma_i\;
  iA \inner \mathcal{P}\of{\sigma_{i+1}}
  \cdots  
  \nonumber\\
  &=&
  \cdots
  \left( 
  \left(A\inner\mathcal{P}A\inner \Gamma\right)\of{\sigma_{i-1}}
  iA\inner \mathcal{P}\of{\sigma_{i+1}}
  -
  iA\inner \mathcal{P}\of{\sigma_{i-1}}
  \left(A\inner \Gamma  A\inner \mathcal{P}\right)\of{\sigma_{i+1}}
  \right)
  \cdots
  \,.
\end{eqnarray}
(This is really a special case of the general formula (3.8) in
\cite{Schreiber:2004e}.)
This way the term
\begin{eqnarray}
  \label{yet another equation}
  iA_\mu\Gamma^\mu\, iA_\nu \mathcal{P}^\nu - iA_\mu\mathcal{P}^\nu\,iA_\mu\Gamma^\mu
  &=&
  \commutator{A_\mu}{A_\nu}\Gamma^\nu\mathcal{P}^\mu
\end{eqnarray}
is produced, and it cancels precisely with 
\refer{another equation}.

This verifies that there are no terms of grade 1.

Now consider the remaining terms of grade 3. It is helpful to write
\begin{eqnarray}
  \label{helpful decopmposition}
  \commutator{A_\mu}{A_\nu}
  \frac{k\inner\Gamma \Gamma^\nu}{2 k\inner \Gamma}
   \mathcal{P^\mu}
  &=&
  \frac{1}{2}
  \commutator{A_+}{A_i} \Gamma^+ \Gamma^i 
  +
  \commutator{A_i}{A_j}
  \frac{k\inner\Gamma \Gamma^j}{2 k\inner \mathcal{P}}
   \mathcal{P}^i
  \,.  
\end{eqnarray}
The first term on the right hand side gives nothing of grade 3 when Poisson-commuted with 
$G_0$. The second term however gives rise to
\begin{eqnarray}
  \label{some intermediate caclulation}
  \commutator{G_0}{  \commutator{A_i}{A_j}\frac{k\inner\Gamma\, \Gamma^j}{2 k\inner \mathcal{P}}
   \mathcal{P}^i
}
  &=&
   \commutator{A_i}{A_j}
    \left(
    \frac{k\inner\Gamma\, \Gamma^j}{2 k\inner \mathcal{P}}
   \Gamma^{\prime i}
  -
    \frac{k\inner\Gamma\, \Gamma^j}{2 (k\inner \mathcal{P})^2}
   k\inner \Gamma^{\prime}
  \right)
  +
  \mbox{terms already considered}
  \nonumber\\
  &=&
   \commutator{A_i}{A_j}
    \left(
    \frac{k\inner\Gamma}{4 k\inner \mathcal{P}}
   (\Gamma^j\Gamma^{i})^\prime
  -
    \frac{k\inner\Gamma\, \Gamma^j}{2 (k\inner \mathcal{P})^2}
   k\inner \Gamma^{\prime}
  \right)
  +
  \mbox{terms already considered}  
  \nonumber\\
  &=&
   \commutator{A_i}{A_j}
    \left(
    \frac{k\inner\Gamma}{4 k\inner \mathcal{P}}
   \Gamma^j\Gamma^{i}
   \right)^\prime
  +
  \alpha
  +
  \mbox{terms already considered}
  \,,
  \nonumber\\  
\end{eqnarray}
where we have abbreviated with
\begin{eqnarray}
  \label{remaining terms}
  \alpha &\defas&
   -
   \commutator{A_i}{A_j}
   \left(
    \left(\frac{k\inner\Gamma}{2 k\inner \mathcal{P}}\right)^\prime
   \Gamma^j\Gamma^{i}   
    +
    \frac{k\inner\Gamma\, \Gamma^j}{2 (k\inner \mathcal{P})^2}
   k\inner \Gamma^{\prime}
  \right)
\end{eqnarray}
two terms which will \emph{not} cancel with anything in the following. (Notice that they are proportional 
to $\sigma$-derivatives of longitudinal objects (along $k$).)

The remaining first term on the right hand side of \refer{some intermediate caclulation}
coalesces with $i A_+ \mathcal{P}^+$ to yield
$\frac{i}{4}\commutator{A_+}{\commutator{A_i}{A_j}}\Gamma^+ \Gamma^i\Gamma^j$.
This cancels against the 
coalescence of \refer{an equation} with the first term on the right hand side of
\refer{helpful decopmposition}
which gives the term
$\frac{i}{2}\commutator{A_j}{\commutator{A_+}{A_i}}\Gamma^j \Gamma^+ \Gamma^i$, because
together they become the longitudinal component of the exterior covariant derivative of the
field strength of $A$, which vanishes. The transversal component of this exterior derivative
of the field strength appears in the remaining terms: 

First there is the remaining coalescence of \refer{an equation} with the second term
on the right hand side of \refer{helpful decopmposition}, which yields
$
  i\commutator{A_k}{\commutator{A_i}{A_j}}
  \frac{k\inner\Gamma}{2k\inner \mathcal{P}}
  \Gamma^k \Gamma^j \mathcal{P}^i
$. Together with the remaining coalescence of the first term on the right of
\refer{some intermediate caclulation} with the transversal $i A_j \mathcal{P}^j$ which
gives rise to
$
  i\commutator{A_k}{\commutator{A_i}{A_j}}
  \frac{k\inner \Gamma}{4 k\inner \mathcal{P}}
  \Gamma^i \Gamma^j\mathcal{P}^k
$
one gets something proportional to
$
  \Big[
    G_0,
  \underbrace{
  \commutator{A_k}{\commutator{A_i}{A_j}}\Gamma^k\Gamma^i\Gamma^j
  }_{=0}
  \Big]
  =0\,,
$
which vanishes because it involves the transversal part of the gauge covariant exterior
derivative of the field strength of $A$.

In summary, the only terms that remain are those of \refer{remaining terms}. When the $\sigma$-derivative
is written out this are three terms which have to vanish seperately, because they
contain different combinations of fermions. Clearly they vanish precisely if \refer{first condition for invariance}
or \refer{second condition for invariance} are satisfied. This completes the proof. \endofproof.

\subsection{Quantum super-Pohlmeyer invariants}
\label{Quantum super-Pohlmeyer invariants}

The DDF-invariants \refer{bosonic susy DDF oscillator} 
are, as discussed in equation (2.12) of \cite{Schreiber:2004b}, still invariants after quantization
in terms of DDF oscillators.
If we take the liberty to denote the quantized objects $\mathcal{P}$ and $\Gamma$ by the same symbols
as their classical counterparts, then the only thing that changes in the notation of the above sections is
that the canonical super-commutation relations \refer{classical CCM} pick up an imaginary factor
\begin{eqnarray}
  \commutator{\mathcal{P}^\mu\of{\sigma}}{\mathcal{P}} &=& -i \eta^{\mu\nu}\delta^\prime\of{\sigma-\kappa}
  \,.
\end{eqnarray}
This again introduces that same factor in the second term of \refer{bosonic susy DDF oscillator} and similarly
in the following expressions.

The quantization of the super-Pohlmeyer invariant \refer{generalized Wilso line} is a trivial consequence
of the quantization of the DDF invariants that it is built from, and, with that imaginary unit taken care of,
its restriction \refer{super-Pohlmeyer invariant} to the case where $R$ is invertible reads
\begin{eqnarray}
 \label{quantized restricted super-Pohlmeyer}
  \mathrm{Tr}\,\mathrm{P}
  \exp\of{
    \int\limits_0^{2\pi}
    d\sigma\;
    \left(
      i A_\mu + \frac{i}{2}\commutator{A_\mu}{A_\nu}\frac{k\inner \Gamma \Gamma^\nu}{k \inner \mathcal{P}}
    \right)
    \mathcal{P}^\mu
  }
  \,.
\end{eqnarray}
Noting that our $A$ is taken to be hermitian and that hence the gauge field strength is
\begin{eqnarray}
  F_A &=& -i(d + iA)^2
  \nonumber\\
  &=&
  dA + i A \wedge A
  \nonumber\\
  &=&
  \left(\partial_{[\mu} A_{\nu]} + \frac{i}{2}\commutator{A_\mu}{A_\nu}\right)dx^\mu \wedge dx^\nu
  \nonumber\\
  &=&
  \frac{1}{2}(F_A)_{\mu\nu}dx^\mu \wedge dx^\nu
\end{eqnarray}
the second term in the integrand is related to the field strength as in
\begin{eqnarray}
  \cdots 
  &=&
  \mathrm{Tr}\,\mathrm{P}
  \exp\of{
    \int\limits_0^{2\pi}
    d\sigma\;
    \left(
      i A_\mu + \frac{1}{2}(F_A)_{\mu\nu}\frac{k\inner \Gamma \Gamma^\nu}{k \inner \mathcal{P}}
    \right)
    \mathcal{P}^\mu
  }
  \,.
\end{eqnarray}
In the case $\commutator{A_i}{A_j} = 0$ \refer{transversal A mutually commute} we hence obtain the
quantized version of \refer{nonlocal susy extension of Pohlmeyer} in the form
\begin{eqnarray}
  \label{restricted and extended superPohl}
  Y[A] 
  &=&
  \mathrm{Tr}\,\mathrm{P}
  \exp\of{
    \int_0^{2\pi}
     d\sigma\;
     \left(
        i A_\mu \mathcal{P}^\mu + \frac{1}{4}(F_A)_{\mu\mu}\Gamma^\mu\Gamma^\nu
     \right)
  }
  \,.
\end{eqnarray}

While the quantized super-Pohlmeyer invariant, being constructed from invariant DDF operators,
is itself a quantum invariant in that it commutes with all the super-Virasoro constraints, the proof
in \S\fullref{Invariance of the extension of the restricted super-Pohlmeyer invariants} of the invariance 
condition of the restricted and then extended form \refer{super-Pohlmeyer invariant} receives quantum corrections. 
In its
classical version the proof makes use of the Grassmann property of the fermions $\Gamma$. Quantumly
there will be diverging contractions in products of $\Gamma$s which not only prevent the application of
the proof to the quantum theory but also make the expression \refer{quantized restricted super-Pohlmeyer}
ill defined without some regularization prescription. As we discuss in the conclusion \S\fullref{Summary and Conclusion}
these regularizations can be done,
but instead of applying them here
we will make contact to the approach which was pioneered in \cite{Hashimoto:2000,Hashimoto:1999} and
generalized to the nonabelian case in \cite{Schreiber:2004e}, where generalized Wilson lines as 
above are applied without regularization to boundary states and the vanishing of divergences in the result 
is then shown to be equivalent
to the equations of motion of the background fields.

This application of \refer{restricted and extended superPohl} to a bare boundary states 
is the content of \S\fullref{Super-Pohlmeyer and boundary states}. But before coming to that
a technicality needs to be discussed, which is done in the next section.

\subsection{On an operator ordering issue in Wilson lines along the closed string}
\label{On an operator ordering issue in Wilson lines along the closed string}

For applying a generalized Wilson line of the kind discussed above to any string state, it 
is helpful to understand how the operators in the Wilson line can be commuted past each other 
to act on the state on the right. It turns out that under a certain condition, which is
fulfilled in the cases we are interested in, the operators can be freely commuted. This works
as follows:\\

A generalized Wilson line of the form
\begin{eqnarray}
  W^{\mathcal{P}}[A] 
   &=& 
  \mathrm{Tr}\,\mathrm{P}\exp\of{\int\limits_0^{2\pi} A\inner \mathcal{P}\of{\sigma}\, d\sigma}
\end{eqnarray}
with even graded $\mathcal{P}$ 
(which could be the $\mathcal{P}$ or $\mathcal{P}^R$ of the
previous sections but also more general objects) breaks up like
\begin{eqnarray}
  W^{\mathcal{P}}[A]
  &=&
  \sum\limits_{n=0}^\infty
  Z^{\mu_1 \cdots \mu_n}
  \mathrm{Tr}\of{A_{\mu_1}\cdots A_{\mu_2}}
\end{eqnarray}
into iterated integrals
\begin{eqnarray}
  \label{definition of Z}
  &&
  \!\!\!\!\!\!\!\!\!\!Z^{\mu_1 \cdots \mu_N}
  \nonumber\\
  &=&
  \frac{1}{N}
  \left[
    \int\limits_{0 < \sigma^1 < \sigma^2 < \cdots < \sigma^N < 2\pi}
    \!\!\!\!\!\!\!\!\!\!\!\!\!\!\!\!\!\!\!\!\!\!d^N \sigma
    \;\;\;\;\;\;\;+ 
    \int\limits_{0 < \sigma^N < \sigma^1 < \cdots < \sigma^{N-1} < 2\pi}
    \!\!\!\!\!\!\!\!\!\!\!\!\!\!\!\!\!\!\!\!\!\!d^N \sigma
    \;\;\;\;\;\;\;+
    \int\limits_{0 < \sigma^{N-1} < \sigma^N < \cdots < \sigma^{N-2} < 2\pi}
    \!\!\!\!\!\!\!\!\!\!\!\!\!\!\!\!\!\!\!\!\!\!d^N \sigma
  \;\;\;\;\;\;\;
  + \cdots\right]  
  \mathcal{P}^{\mu_1}\of{\sigma^1}
  \cdots
  \mathcal{P}^{\mu_N}\of{\sigma^N}
  \,.
  \nonumber\\
\end{eqnarray}
In equation (2.17) of \cite{Schreiber:2004b} if was noted that the integration domain
can equivalently be written as
\begin{eqnarray} 
  \label{alternative version of Z}
  Z^{\mu_1 \cdots \mu_N}
  &=&
  \frac{1}{N}
  \int\limits_0^{2\pi}
  d\sigma^1
  \;
  \int\limits_{\sigma^1}^{\sigma^1 + 2\pi}
  d\sigma^2
  \;
  \cdots
  \int\limits_{\sigma^{N-1}}^{\sigma^1 + 2\pi}
  d\sigma^N
  \;  
  \mathcal{P}^{\mu_1}\of{\sigma^1}
  \mathcal{P}^{\mu_2}\of{\sigma^2}
  \cdots
  \mathcal{P}^{\mu_N}\of{\sigma^N}
  \,.
\end{eqnarray}
This is seen by simply replacing all $\sigma^i < \sigma^1$ for $i>1$ by $\sigma^i + 2\pi$. Due to
the periodicity of $\mathcal{P}$ this does not change the value of the integral but yields the
integration bounds used in \refer{alternative version of Z}.

The reason why this is recalled here is that a slight generalization of this fact will be needed
in the following. Namely for any integer $M$ with $1 < M < N$ one can obviously more generally write
\begin{eqnarray}
  \label{yet another version of Z}
  &&Z^{\mu_1 \cdots \mu_N}
  \nonumber\\
  &=& 
  \int\limits_0^{2\pi}
  d\sigma^1
  \;
  \int\limits_{\sigma^1}^{\sigma^1 + 2\pi}
  d\sigma^M
  \;\;
  \int\limits_{\sigma^1}^{\sigma^M}
  d\sigma^2
  \;
  \cdots
  \int\limits_{\sigma^{M-2}}^{\sigma^M}
  d\sigma^{M-1}
  \;\;
  \int\limits_{\sigma^{M}}^{\sigma^1 + 2\pi}
  d\sigma^{M+1}
  \cdots
  \int\limits_{\sigma^{N-1}}^{\sigma^1 + 2\pi}
  d\sigma^N
  \;  
  \mathcal{P}^{\mu_1}\of{\sigma^1}
  \mathcal{P}^{\mu_2}\of{\sigma^2}
  \cdots
  \mathcal{P}^{\mu_N}\of{\sigma^N}
  \,.
  \nonumber\\
\end{eqnarray}
Equation \refer{alternative version of Z} follows as the special case with $M=2$.

The motivation for these considerations is the following:

Classically, the $\mathcal{P}$ commute among each other. Therefore the ordering of the
$\mathcal{P}$ in the integrand makes no difference, only the combination of spacetime
index $\mu_i$ with integration variable $\sigma^i$ does.

Here we want to note that this remains true at the quantum level \emph{if}
\begin{eqnarray}
  \label{condition for interchangability of objects in Wilson line}
  \commutator{\mathcal{P}\of{\sigma}}{\mathcal{P}\of{\kappa}} 
  &\propto& 
  \delta^\prime\of{\sigma - \kappa}
  \,.
\end{eqnarray}

This is readily seen by commuting $\mathcal{P}\of{\sigma^1}$ with $\mathcal{P}\of{\sigma^M}$
in \refer{yet another version of Z}. The result has the form
\begin{eqnarray}
  \int\limits_0^{2\pi}
  d\sigma^1
  \;
  \int\limits_{\sigma^1}^{\sigma^1 + 2\pi}
  d\sigma^M
  \;
  \delta^\prime\of{\sigma^1 - \sigma^M}
  F\of{\sigma^1, \cdots, \sigma^N}
  &=&
  \int\limits_0^{2\pi}
  d\sigma^1
  \;
  \int\limits_{\sigma^1}^{\sigma^1 + 2\pi}
  d\sigma^M
  \;
  \delta\of{\sigma^1 - \sigma^\kappa}
  \frac{\partial}{\partial \sigma^M}
  F\of{\sigma^1, \cdots, \sigma^N}
  \nonumber\\
  &=&
  \int\limits_{\sigma^1}^{\sigma^1 + 2\pi}
  d\sigma^M
  \;
  \delta\of{\sigma^1 - \sigma^\kappa}
  \frac{\partial}{\partial \sigma^M}
  F\of{\sigma^1, \cdots, \sigma^N}
  \nonumber\\
  &=&
  0
  \,,
\end{eqnarray}
so that all resulting commutator terms vanish. Every other commutator can be obtained by 
using the cyclic invariance in the integration variables.

More generally, any two (even graded, periodic) 
objects $A\of{\sigma}$, $B\of{\kappa}$ in the integrand of an iterated 
integral of the form \refer{definition of Z} whose commutator is proportional to 
$\commutator{A\of{\sigma}}{B\of{\kappa}} \propto \delta^\prime\of{\sigma-\kappa}$ can be 
commuted past each other in the Wilson line without affecting the value of the integral.

This simple but crucial observation will be needed below for the demonstration that
Pohlmeyer-invariants map the boundary state of a bare D-brane to that describing a brane
with a nonabelian gauge field turned on.

\subsection{Super-Pohlmeyer and boundary states}
\label{Super-Pohlmeyer and boundary states}

We now have all ingredients in place to apply the super-Pohlmeyer invariant to the boundary state
of a bare D9 brane. A brief review of the idea of boundary states adapted to the present context
is given in \cite{Schreiber:2004e}, but in fact only two simple relations are needed for the following:

If $\ket{\mathrm{D9}}$ is the boundary state of the space-filling BPS D9 brane, then
(Due to equation (2.26) in \cite{Schreiber:2004b} and section 2.3.1 in \cite{Schreiber:2004e})
we have
\begin{eqnarray}
  \mathcal{P}^\mu\of{\sigma}\ket{\mathrm{D9}}
  &=&
  \sqrt{
    \frac{T}{2}
  }
    X^{\prime\mu}\of{\sigma}\ket{\mathrm{D9}}
\end{eqnarray}
and
\begin{eqnarray}
  \Gamma^\mu\of{\sigma}
  \ket{\mathrm{D9}}
  &=&
  {\cal E}^{\dagger \mu}\of{\sigma}
  \ket{\mathrm{D9}}
  \,.
\end{eqnarray}

Using the results of \S\fullref{On an operator ordering issue in Wilson lines along the closed string}
such a replacement extends to the full Wilson line made up from these objects:

Consider the extension \refer{restricted and extended superPohl} of the restricted super-Pohlmeyer invariant
with $A_+ \neq 0$ and furthermore only mutually commuting spatial components of $A$ nonvanishing. In this case
the fermionic terms in the integrand have trivial commutators so that the integrand as a whole 
satisfies condition \refer{condition for interchangability of objects in Wilson line}. Therefore,
according to the result of \S\fullref{On an operator ordering issue in Wilson lines along the closed string},
we can move all appearances of $\mathcal{P}^\mu  + \frac{1}{4}(F_A)_{\mu\nu}\Gamma^\mu\Gamma^\nu$
to the boundary state $\ket{\mathrm{D9}}$ on the right, change it there to 
$\sqrt{\frac{T}{2}}X^{\prime \mu} + \frac{1}{4}(F_A)_{\mu\nu}\mathcal{E}^{\dagger \mu}\mathcal{E}^{\dagger \nu}$
and then move this back to the original position (noting that still 
$\commutator{X^{\prime}\of{\sigma}}{\mathcal{P}\of{\kappa}} \propto \delta^\prime\of{\sigma-\kappa}$).
This way we have
\begin{eqnarray}
  \label{from Pohlmeyer to boundary}
  &&\mathrm{Tr}\, \mathrm{P}
  \exp\of{
    \int\limits_0^{2\pi}
     d\sigma\,
    \left(
      i A_\mu \mathcal{P}^\mu + \frac{1}{4}(F_A)_{\mu\nu}\Gamma^\mu\Gamma^\nu
    \right)
  }
  \ket{\mathrm{D9}}
  \nonumber\\
  &=&
  \mathrm{Tr}\, \mathrm{P}
  \exp\of{
    \int\limits_0^{2\pi}
     d\sigma\,
    \left(
      -i 
      \sqrt{\frac{T}{2}} A_\mu X^{\prime \mu} + \frac{1}{4}(F_A)_{\mu\nu}
      \mathcal{E}^{\dagger \mu}
      \mathcal{E}^{\dagger \nu}
    \right)
  }
  \ket{\mathrm{D9}}
  \,.
\end{eqnarray}

If we allowed ourself to regulate all the generalized Wilson lines considered here by a point-splitting
method as in \cite{MaedaNakatsuOonishi:2004}, i.e. by taking care that no local fields in the
Wilson line ever come closer than some samll distance $\sigma$, then the above step becomes a triviality.
Indeed, the result of \cite{MaedaNakatsuOonishi:2004} together with those of
\cite{Hashimoto:2000,Hashimoto:1999,Schreiber:2004e}
shows that this is a viable approach, because the condition for the $\epsilon$-regularized
Wilson line to be still an invariant is the same as that of the non-regularized Wilson line to be
free of divergences and hence well defined. This is discussed further in 
\S\fullref{Summary and Conclusion}.

It will be convenient for our purposes to rescale $A$ as
\begin{eqnarray}
  A \mapsto -\sqrt{\frac{2}{T}}A
  \,, 
\end{eqnarray}
so that this becomes
\begin{eqnarray}
  \label{boundary state from super-Pohlmeyer}
  \cdots 
  &=&
  \mathrm{Tr}\, \mathrm{P}
  \exp\of{
    \int\limits_0^{2\pi}
     d\sigma\,
    \left(
      i 
      A_\mu X^{\prime \mu} 
      + 
      \frac{1}{2T}(F_A)_{\mu\nu}
      \mathcal{E}^{\dagger \mu}
      \mathcal{E}^{\dagger \nu}
    \right)
  }
  \ket{\mathrm{D9}}
  \,.
\end{eqnarray}
This is finally our main result, because this is precisely the boundary state of a nonabelian gauge field
as considered in 
equation (3.14) of \cite{Schreiber:2004e}, which is a generalization of the abelian case studied
in \cite{Hashimoto:2000,Hashimoto:1999}. The same form of the boundary state is obtained from
equations (3.7), (3.8)
in \cite{MaedaNakatsuOonishi:2004} 
when in the expression given there the integral over the Grassmann variables is performed
(following the computation described on pp. 236-237 of \cite{AndreevTseytlin:1988}).

The boundary state \refer{boundary state from super-Pohlmeyer} has two important properties:

\begin{enumerate}

\item

\paragraph{Super-Ishibashi property of the boundary state.}

The defining property of boundary states is that they are annihilated by the generators 
$\mathcal{L}_K$ of
$\sigma$-reparameterization as well as, in the superstring case, by their square root 
$\extd_K$, which is a deformed exterior derivative on loop space. $\mathcal{L}_K$ is a
linear combination of left- and right-moving bosonic super-Virasoro generators, while 
$\extd_K$ is a combination of fermionic super-Virasoro generators, as discussed in 
\cite{Schreiber:2004e}. 

It is noteworthy that the state \refer{boundary state from super-Pohlmeyer} 
indeed satisfies the Ishibashi conditions.
Naively this must be the case, because this state is obtained from the bare $\ket{\mathrm{D9}}$,
which does satisfy it by definition, by acting on it with a super-Pohlmeyer operator, that
commutes with all constraints and hence leaves the Ishibashi property of $\ket{\mathrm{D9}}$
intact. But above we mentioned that the restricted form \refer{quantized restricted super-Pohlmeyer}
of the quantized Pohlmeyer invariants that this state comes from has potential quantum anomalies
which would spoil this invariance. These are due to the non-Grassmann property of the
quantized fermions $\Gamma$. However, after application to the bare $\ket{\mathrm{D9}}$
which gives \refer{boundary state from super-Pohlmeyer}, the left- and right-moving 
fermions are replaced by their polar combination $\mathcal{E}^\dagger$, and these again
enjoy the Grassmann property (they are nothing but differential forms on loop space). For this
reason the final result can again enjoy the Ishibashi property, which means nothing but
super-reparameterization invariance with respect to $\sigma$.

{\it Proof:}

The invariance under reparameterizations induced by $\mathcal{L}_K$ is manifest, analogous
in all the cases considered here before, since \refer{boundary state from super-Pohlmeyer} is
the generalized Wilson line over an object of unit reparameterization weight. 

The only nontrivial
part that hence needs to be checked is the commutation with $\extd_K $ and here we only need to
know that
$\commutator{\extd_K}{X^\mu\of{\sigma}} = \mathcal{E}^{\dagger \mu}\of{\sigma}$.

Applying this to \refer{boundary state from super-Pohlmeyer}
we get, in the same manner as in the similar computations before,
from $\commutator{\extd_K}{i A_\mu X^{\prime \mu}} =
 i (\partial_\mu A_\nu - \partial_\nu A_\mu)\mathcal{E}^{\dagger \mu}X^{\prime \nu}
+ \left(iA_\mu \mathcal{E}^{\dagger \mu}\right)^\prime$
coalesced terms
$-\commutator{A_\mu}{A_\nu}\mathcal{E}^{\dagger \mu}X^{\prime \nu}$
and
$\frac{i}{2T}\commutator{A_\kappa}{(F_A)_{\mu\nu}}\mathcal{E}^{\dagger \kappa}\mathcal{E}^{\dagger \mu}\mathcal{E}^{\dagger \nu}$
at the integration boundaries.

These combine with the terms
$\commutator{\extd_K}{\frac{1}{2T}(F_A)_{\mu\nu}\mathcal{E}^{\dagger \mu}\mathcal{E}^{\dagger \nu}}
=
\frac{1}{2T} (\partial_{[\kappa} (F_A)_{\mu\nu]}
  \mathcal{E}^{\dagger \kappa}\mathcal{E}^{\dagger \mu}\mathcal{E}^{\dagger \nu})
+
-i (F_A)_{\mu\nu}\mathcal{E}^{\dagger \mu}X^{\prime \nu}
$
to
$
  \left(
   i (\partial_\mu A_\nu - \partial_\nu A_\mu) - \commutator{A_\mu}{A_\nu} -i (F_A)_{\mu\nu}
  \right)\mathcal{E}^{\dagger \mu}X^{\prime \nu}
  = 0
$
and\\
$
 \frac{1}{2T}
  \left(
    \partial_{[\kappa}(F_A)_{\mu\nu}
    +
    i \commutator{A_{[\kappa}}{(F_A)_{\mu\nu]}}
  \right)
  \mathcal{E}^{\dagger \kappa}\mathcal{E}^{\dagger \mu}\mathcal{E}^{\dagger \nu}
  = 0
  \,.
$
Hence all terms vanish and $\extd_K$ commutes with \refer{boundary state from super-Pohlmeyer}. \endofproof

\item

\paragraph{Nonlinear gauge invariance of the boundary state.}

A generic state constructed from gluon vertices for nonabelian $A$ will generically not
be invariant under a target space gauge transformation $A \to UAU^\dagger + U(dU^\dagger)$.
The generalized Wilson line in \refer{boundary state from super-Pohlmeyer} however does
have this invariance  - at least at the classical level. This follows from the general
invariance properties of Wilson lines (for details see appendix B of \cite{Schreiber:2004e})
and depends crucially on the appearance of the gauge covariant field strength $F_A$ in 
\refer{boundary state from super-Pohlmeyer}. 

\end{enumerate}

Further comments on the nature of the boundary states considered here,
in particular on the question of divergences and their regularization, are left to the concluding section.

\section{Conclusion}
\label{Summary and Conclusion}

The super-Pohlmeyer invariants, whose construction principle was only briefly mentioned
at the end of \cite{Schreiber:2004b}, have been studied here in more detail. In particular
their expression in terms of local fields has been worked out on that part of phase space
where it exists. The extension of this expression to the full  phase space has been found to be
invariant on all of phase space if the transversal components of the gauge field mutually commute.

Quantizing the result and applying it to the boundary state of a bare D9 brane was shown to yield
boundary states of the form considered before in 
\cite{MaedaNakatsuOonishi:2004,Schreiber:2004e}, which are straightforward
non-abelian generalizations of those studied in \cite{Hashimoto:2000}.

This result shows that the Pohlmeyer invariants, which before have only appeared in the 
connection with vague hopes to circumvent well-known quantum effects like the critical dimension,
do have a role to play in (standard) string theory. 

While this might not appear as much of a surprise in light of the result
\cite{Schreiber:2004b} which showed that the Pohlmeyer invariants are a subset of all
DDF invariants that have the crucial property of mapping physical states to physical states
by commuting with all the (super-)Virasoro constraints, it seems noteworthy that 
it is a priori not obvious that this ability to generate the physical spectrum 
translates also to a generation of boundary states that satisfy the Ishibashi condition.

And indeed, it was shown above that a precise match between super-Pohlmeyer invariants
and boundary states is manifest only in the special case where the transversal components
of the gauge field mutually commute. On the other hand, we did not show that in other cases
the result of acting with the super-Pohlmeyer invariant on the bare D9-brane boundary state
does not produce a new some boundary, but it is at least not obvious.

Of course, naively it is obvious that the result of acting with a super-Pohlmeyer invariant
on a state which satisfies the Ishibashi conditions still satisfies these conditions,
simply due to the very invariance property of the Pohlmeyer invariants. But the result will
in general need regularization and/or a condition on the gauge field that ensures vanishing
of contact term divergences, and for the case where the transversal components of $A$ do not mutually
commute we did not show that the result has this property.

This is in contrast to the case where the transversal components do mutually commute, and where
the lightlike component of $A$ which is non-parallel to $k$ vanishes. In this case we did show that the
result of acting with the super-Pohlmeyer invariant associated with that $A$-field on 
$\ket{D9}$ does produce the result \refer{boundary state from super-Pohlmeyer}. And this result
is known to be well defined if (and only if) $A$ satisfies its background equations of motion.

This can either be seen by introducing a regularization and checking that this regularization preserves
the invariance properties iff the background equations of motion hold, as was done in
\cite{MaedaNakatsuOonishi:2004}. Or, alternatively, no regularization is used and the resulting
divergences are shown to vanish when $A$ satisfies its equations of motion. This was done
for the abelian case to second order in \cite{Hashimoto:2000,Hashimoto:1999} and for the
non-abelian case to first order in \cite{Schreiber:2004e}.

While this relation between quantum divergences and background equations of motion is perfectly
natural in the context of boundary states, it is a new aspect in the study of Pohlmeyer
invariants. It shows that even though these have a consistent quantization in terms of 
DDF invariants in the sense of constituting a closed algebra of quantum operators that commute
with the constraints, not all of them have a well defined application on all states of the
string's Hilbert space. Namely, even though they consist of well-behaved DDF operators, these
appear in infinite sums and divergences may occur when acting with these on some states.

\acknowledgments{
I am grateful to Robert Graham for his comments on the ideas presented here as well as to
Rainald Flume for interesting discussion about Pohlmeyer invariants while these ideas were being
developed.

This work was supported by SFB/TR 12.
}

\newpage

\appendix

\newpage

\bibliography{std}

\end{document}